\documentstyle[amssymb,aps,preprint,12pt]{revtex}

\begin{document}
\title{Effective Widths and Effective Number of Phonons of Multiphonon Giant
Resonances$^{\thanks{%
Supported in part by CNPq and FAPESP.}}$}
\author{L.F. Canto$^{1}$, B.V.~Carlson$^{2}$, M.S.~Hussein$^{3}$ and A.F.R. de
Toledo Piza$^{3}$}
\address{$^{1}$Instituto de F\'{\i}sica, Universidade do Rio de Janeiro,\\
CP 68528, 21945-970 Rio de Janeiro RJ, Brazil\\
$^{2}$Departamento de F\'{i}sica, Instituto Tecnol\'{o}gico de
Aeron\'{a}utica - CTA\\
12228-900 S\~ao Jos\'e dos Campos, SP, Brazil\\
$^{3}$Instituto de F\'{\i}sica, Universidade de S\~{a}o Paulo,\\
C.P. 66318, 05315-970 S\~{a}o Paulo, Brazil}
\date{\today}
\maketitle

\begin{abstract}
We discuss the origin of the difference between the harmonic value of the
width of the multiphonon giant resonances and the smaller observed value.
Analytical expressions are derived for both the effective width and the
average cross-section. The contribution of the Brink-Axel mechanism in
resolving the discrepancy is pointed out.
\end{abstract}

\newpage

In a series of publications \onlinecite{ref1,ref2,ref3}, we have advanced
the idea that the excitation of the double giant dipole resonances (DGDR) in
heavy-ion collisions proceeds in two distinct, incoherent, ways: the usual
sequential excitation via the single giant dipole resonance (GDR), usually
referred to in the literature as the ``harmonic'' cross section~%
\onlinecite{ref4}, and the fluctuation excitation that involves the
``internal'' decay of the GDR into the complicated background states
followed by a collective excitation of a GDR on these states, a
manifestation of the Brink-Axel mechanism. Of course in a fully microscopic
approach to the excitation process, the Brink-Axel effects are also
contained. However, the usefulness of the approach of \onlinecite{ref1,ref3}
is the clear identification of the average cross-section with a ``direct''
excitation and the fluctuation one with the higher-order ``non-direct''
contribution. This allows identifying different time scales with the
excitation process which makes possible an unambiguous treatment of the
eventual decay of the multiphonon states [2]. The question of the
``enhancement'' of the experimental cross-section with respect to the
harmonic value is easily addressed within the average$+$fluctuation model %
\onlinecite{ref1,ref3}. However, the enhancement problem, especially in $%
^{136}$Xe still remains not fully resolved. There has also been some
discussion in recent years related to the fact that the observed width of
the double giant dipole resonance deviates from the harmonic value of twice
the single giant resonance width. As we have commented previously,~\cite
{ref1} this is a natural result of the incoherent contributions to the cross
section, even when the resonances themselves are purely harmonic. In fact,
due to the energy dependence of the incoherent contributions to the
excitation cross sections, we expect the effective $n$-phonon width to be
energy dependent. The purpose of this paper is to derive an analytical
expression for the effective energy-dependent width of the n-phonon
excitation distribution. For this purpose we use multistep distorted wave
series in conjunction with statistical averaging that allows us to
significantly extend the result of our previous publication [1]. The method
enables us to calculate the inclusive n-phonon cross-section in closed form
and allows us to identify the physical quantity responsible for the
cross-section enhancement. This quantity is the product of the damping width
of the single phonon resonance times the collision time evaluated at the
grazing impact parameter $b_{o}$.

It is convenient to introduce projection operators: $P$ will stand for the
entrance channel, $d_{1}$ the 1 phonon channel, $q_{1}$ the corresponding
fine structure subspace etc. We use multistep distorted wave picture. Thus
for the excitation of e.g. 2 phonons, one has

\begin{equation}
\sigma ^{\left( 2\right) }=\left| \left\langle x_{f}^{\left( -\right)
},f\left| V\text{ }G_{1}^{\left( +\right) }\text{ }V\right| x_{i}^{\left(
+\right) },0\right\rangle \right| ^{2}\text{ .}
\end{equation}

In the above $V$ stands for the Coulomb interaction responsible for the
excitation, $x^{\left( +\right) }$ stands for the appropriate distorted wave
and $G_{1}^{\left( +\right) }$ is the Green's \ operator in the subspace $%
d_{1}+q_{1}$. This is given by

\begin{equation}
G_{1}^{+}=\left( 
\begin{array}{c}
g_{d_{1}d_{1}}^{\left( +\right) } \\ 
G_{q_{1}}^{\left( +\right) }\text{ }v_{q_{1}d_{1}}\text{ }%
g_{d_{1}d_{1}}^{\left( +\right) }
\end{array}
\begin{array}{c}
G_{d_{1}}^{\left( +\right) }\text{ }v_{d_{1}q_{1}}\text{ }%
g_{q_{1}q_{1}}^{\left( +\right) } \\ 
g_{q_{1}q_{1}}^{\left( +\right) }
\end{array}
\right) \text{,}
\end{equation}

\noindent where, $g_{ii}^{\left( +\right) }=\left( E^{\left( +\right)
}-H_{i}-v_{ij}G_{j}^{+}v_{ji}\right) ^{-1}$, $G_{i}^{\left( +\right)
}=\left( E^{\left( +\right) }-H_{i}\right) ^{-1}$, $H_{i}=h_{rel}+h_{i}$,
and $E^{\left( +\right) }=E+i\varepsilon $, with $h_{rel}$ representing the
Hamiltonian of relative motion and $h_{i}$ the nuclear Hamiltonian of ion $i$%
. The interaction $v$, couples the different subspaces.

The interaction $V$ connects the ground state to the one-phonon doorway
state $d_{1}$. It can also connects a state in $q_{1}$to a one-phonon
Brink-Axel state. Thus the amplitude that enters into the definition of the
cross-section $\sigma ^{(2)}$ separates into two pieces

\begin{eqnarray}
\sigma ^{\left( 2\right) } &=&\left| \left\langle x_{f}^{\left( -\right)
},2|V_{d_{2}d_{1}}\left( E^{\left( +\right)
}-H_{d_{1}}-v_{d_{1}q_{1}}G_{q_{1}}v_{q_{1}d_{1}}\right)
^{-1}V_{d_{1},0}\right| x_{i}^{\left( +\right) },0\right\rangle  \nonumber \\
&&+\left\langle x_{f}^{\left( -\right) },1\left| V_{d_{1}q_{1}}\frac{1}{%
E^{\left( +\right) }-H_{q_{1}}}v_{q_{1}d_{1}}\frac{1}{E^{\left( +\right)
}-H_{d_{1}}-v_{d_{1}q_{1}}G_{q_{1}}v_{q_{1}d_{1}}}V_{d_{1},0}|x_{i}^{\left(
+\right) },0\right\rangle \right| ^{2}\text{ .}
\end{eqnarray}

The fine structure states $\{q_{1}\}$ are quite complicated many
particle-many hole configurations. In this respect, a statistical treatment
of the second, fluctuation, contribution is called for. Thus we take the
energy average of $\sigma ^{\left( 2\right) }$ to represent the theoretical
cross-section to be compared to the data. When performing the energy
average, the cross term in Eq. (4) vanishes since it is linear in the
assumed random coupling $v_{q_{1}d_{1}}$. Thus $\overline{\sigma ^{\left(
2\right) }}$ is given, as usual, by the incoherent sum of two terms. The
first term is the coherent ``harmonic'' cross-section. Here the Green's
function $\left( E-H_{d_{1}}-v_{d_{1}q_{1}}G_{q_{1}}v_{q_{1}d_{1}}\right)
^{-1}$ is replaced by an average over the $q_{1}$ states $\left( E-H_{d_{1}}-%
\overline{v_{d_{1}q_{1}}G_{q_{1}}v_{q_{1}d_{1}}}\right) ^{-1}$. The
correction to this approximation involves fluctuations that corresponds to
the process $d_{1}\rightarrow q_{1}\rightarrow d_{1}$, which is negligible.
The average $\overline{v_{d_{1}q_{1}}G_{q_{1}}v_{q_{1}d_{1}}}$ is as usual
written as $\Delta _{d_{1}}-i\frac{\Gamma _{d_{1}}}{2}$, which defines the
resonance shift $\Delta _{d_{1}}$ and width $\Gamma _{d_{1}}$ respectively.
These two quantities satisfy a dispersion relation. The fluctuation
contribution to $\overline{\sigma ^{\left( 2\right) }}$ involves an average
over the $q_{1}$ states.

This average is easily performed if we recall that the $q_{1}$ Green's
function is a large sum of random contributions. Thus, writing first

\begin{eqnarray}
G_{q}^{\left( +\right) } &=&%
\mathrel{\mathop{\sum }\limits_{q_{1}}}%
\frac{\left| q_{1}\right\rangle \left\langle q_{1}\right| }{E^{\left(
+\right) }-\varepsilon _{q_{1}}+i\frac{\Gamma _{0q_{1}}}{2}-h_{rel}} 
\nonumber \\
&\equiv &\int \frac{d\vec{k}}{\left( 2\pi \right) ^{3}}%
\mathrel{\mathop{\sum }\limits_{q_{1}}}%
\frac{\left| x_{k}^{\left( +\right) }\right\rangle \left| q_{1}\right\rangle
\left\langle q_{1}\right| \left\langle x_{k}^{\left( +\right) }\right| }{%
E^{\left( +\right) }-\varepsilon _{q_{1}}+i\frac{\Gamma _{0q_{1}}}{2}-E_{k}}%
\text{ ,}
\end{eqnarray}

\noindent where we have employed the scattering eigenstates of $h_{rel}$, $%
\left\{ \left| x_{k}^{\left( +\right) }\right\rangle \right\} $, we can
reduce the average above into

\noindent 
\begin{eqnarray}
\overline{\sigma _{fl}^{\left( 2\right) }} &=&\int \frac{d\vec{k}}{\left(
2\pi \right) }\int \frac{d\vec{k}^{\prime }}{\left( 2\pi \right) ^{3}}%
\left\langle 0,x_{i}^{\left( +\right) }\left| V^{\dagger }g_{d_{1}}^{\dagger
}\right| d_{1}\right\rangle \overline{%
\mathrel{\mathop{\sum }\limits_{q_{1}}}%
\frac{\left\langle d\left| v^{\dagger }\right| q_{1}x_{k}^{\left( +\right)
}\right\rangle \left\langle q_{1}x_{k}^{\left( +\right) }\left| V^{\dagger
}\right| x_{f}^{\left( -\right) },1\right\rangle }{E-\varepsilon _{q_{1}}-i%
\frac{\Gamma _{0q_{1}}}{2}-E_{k}}}  \nonumber \\
&&\cdot \overline{%
\mathrel{\mathop{\sum }\limits_{q_{1}^{\prime }}}%
\frac{\left\langle x_{f}^{\left( -\right) },1\left| V\right| q\prime
,x_{k\prime }^{\left( +\right) }\right\rangle \left\langle q\prime
,x_{k\prime }^{\left( +\right) }\left| v\right| d_{1}\right\rangle }{%
E-\varepsilon _{q_{1}^{\prime }}+i\frac{\Gamma _{0q_{1}^{\prime }}}{2}%
-E_{k\prime }}}\left\langle d_{1}\left| g_{d}V\right| 0,x_{i}^{\left(
+\right) }\right\rangle \text{ .}
\end{eqnarray}

The average over $\{q_{1}\}$ makes the double $q_{1}$ sum collapses into a
single sum $\left( q_{1}=q_{1}^{\prime }\right) $. Further, since there are
many $q_{1}$ states, we replace the $q_{1}$ sum by an integral $
\begin{array}{l}
\mathrel{\mathop{\sum }\limits_{q_{1}}}%
=\int d\varepsilon _{q_{1}}\rho \left( \varepsilon _{q_{1}}\right)
\end{array}
$, where $\rho \left( \varepsilon _{q_{1}}\right) $ is the density of the $%
q_{1}$ states. The $\varepsilon _{q}$ integral can be easily performed if we
assume that the numerator is slowly varying. Only the two $\varepsilon
_{q}-poles$ will contribute. Thus

\begin{eqnarray}
\overline{\sigma _{fl}^{\left( 2\right) }} &=&\Gamma _{d_{1}}^{\downarrow
}\int \frac{d\vec{k}}{\left( 2\pi \right) ^{3}}\int \frac{d\vec{k}^{\prime }%
}{\left( 2\pi \right) ^{3}}\left\langle 0x_{i}^{\left( +\right) }\left|
V^{\dagger }g_{d_{1}}^{\dagger }\right| d_{1}x_{k}^{\left( +\right)
}\right\rangle \left\langle \tilde{0}\left( \overline{\varepsilon }%
_{q}\right) x_{k}^{\left( +\right) }\left| V^{\dagger }\right|
1,x_{f}^{\left( -\right) }\right\rangle  \nonumber \\
&&\cdot \frac{1}{\left| -\overline{\Gamma }_{0q_{1}}+i\left(
E_{k}-E_{k^{\prime }}\right) \right| }\left\langle x_{f}^{\left( -\right)
},1\left| V\right| \tilde{0}\left( \overline{\varepsilon }_{q}\right)
x_{k^{\prime }}^{\left( +\right) }\right\rangle \left\langle x_{k^{\prime
}}^{\left( +\right) }d_{1}\left| g_{d_{1}}V\right| 0x_{i}^{\left( +\right)
}\right\rangle \text{ ,}
\end{eqnarray}

\noindent where we have introduced the spreading width of the doorway state $%
d_{1}$, defined by

\begin{equation}
\Gamma _{d_{1}}^{\downarrow }=2\pi \rho _{q_{1}}\overline{\left| \langle
q_{1}\right| v\left| d_{1}\rangle \right| ^{2}}\text{ ,}
\end{equation}

\noindent and also introduced a representative excited intrinsic state $%
\left| \tilde{0}\left( \overline{\varepsilon }_{q}\right) \right\rangle $,
which acts as a ground state for the Brink-Axel phonon excitation.

The energy correlation function $\left| -\Gamma _{0q_{1}}+i\left(
\varepsilon _{k}-\varepsilon _{k^{\prime }}\right) \right| ^{-1}$ controls
the magnitude of $\sigma _{fl}^{\left( 2\right) }$. For $\varepsilon
_{k}=\varepsilon _{k^{\prime }}$, and $\Gamma _{0q_{1}}\rightarrow 0$, the
contribution is much smaller than $\sigma _{h}^{\left( 2\right) }$ since
only one $\vec{k}$ integral survive in the former. On the other hand, if $%
\overline{\Gamma }_{0q_{1}}$ is much larger than the range of values of $%
\varepsilon _{k}-\varepsilon _{k^{\prime }}$ which an relevant for the
integral, then $\overline{\sigma }_{fl}^{\left( 2\right) }$ would be $\frac{%
\Gamma _{d_{1}}^{\downarrow }}{\overline{\Gamma }_{0q_{1}}}$ times a regular
cross section. However, $\overline{\Gamma }_{0q_{1}}$ is by definition $\ll
\Gamma _{d}^{\downarrow }$, and therefore this last case does not occur.
Therefore we may simply say that the correlation function introduces a
characteristic time, the \ ``correlation'' time, $\tau _{c}$ which depends
on the bombarding energy. Accordingly we write for $\overline{\sigma
_{fl}^{\left( 2\right) }}$ the following

\begin{equation}
\overline{\sigma _{fl}^{\left( 2\right) }}=\Gamma _{d_{1}}^{\downarrow }\tau
_{c}\left( E\right) \left| \left\langle x_{f}^{\left( -\right) },1\left|
V\right| \tilde{0}\left( \varepsilon _{d_{1}}\right) \right\rangle
\left\langle d_{1}|g_{d}V|0x_{i}^{\left( +\right) }\right\rangle \right| ^{2}%
\text{ ,}
\end{equation}

\noindent where the integrals $\int \frac{d\vec{k}^{\prime }}{\left( 2\pi
\right) ^{3}}\int \frac{d\vec{k}}{\left( 2\pi \right) ^{3}}\left|
x_{k^{\prime }}^{\left( +\right) }\right\rangle \left| x_{k}^{\left(
+\right) }\right\rangle \left\langle x_{k}^{\left( +\right) }\right|
\left\langle x_{k^{\prime }}^{\left( +\right) }\right| $ have been set equal
to unity since the distorted waves sts $\{x^{\left( +\right) }\}$ is
complete.

Now since $\left\langle x_{f}^{\left( -\right) },1\left| V\right| \tilde{0}%
\left( \varepsilon _{d_{1}}\right) \right\rangle $, contains one Brink-Axel
phonons in the final state, where as the corresponding amplitude in the
harmonic cross-section, $\left\langle x_{f}^{\left( -\right) },2\left|
V\right| 1\right\rangle $, contains 2, we can make the approximation $\frac{%
\left\langle x_{f}^{\left( -\right) },1\left| V\right| \tilde{O}%
\right\rangle }{\left\langle x_{f}^{\left( -\right) },2\left| V\right|
1\right\rangle }\cong \frac{1}{\sqrt{2!}}$.

Thus we find for $\overline{\sigma _{fl}^{\left( 2\right) }}$, the following
reasonable approximate form

\begin{equation}
\overline{\sigma _{fl}^{\left( 2\right) }}=\frac{1}{2}\frac{\Gamma
_{d_{1}}^{\downarrow }\tau _{c}\left( E\right) }{\hslash }\sigma
_{c}^{\left( 2\right) }\text{ .}
\end{equation}

\noindent Thus

\begin{equation}
\overline{\sigma ^{\left( 2\right) }}=\sigma _{c}^{\left( 2\right) }\left( 1+%
\frac{1}{2}\frac{\Gamma _{d_{1}}^{\downarrow }\tau _{c}\left( E\right) }{%
\hslash }\right) \text{ .}
\end{equation}

The second term above gradually becomes insignificant as the bombarding
energy increases.

The calculation of the cross-section for higher number of phonons follows a
similar procedure. Keeping track of the number of routes that can be
followed to reach the final \ ``state'', we have for $\sigma ^{\left(
3\right) }$

\begin{equation}
\overline{\sigma ^{\left( 3\right) }}=\sigma _{c}^{\left( 3\right) }\left( 1+%
\frac{2}{3}\frac{\Gamma _{d_{1}}^{\downarrow }\tau _{c}\left( E\right) }{%
\hslash }+\frac{1}{3}\left( \frac{\Gamma _{d_{1}}^{\downarrow }\tau
_{c}\left( E\right) }{\hslash }\right) ^{2}\right) \text{ .}
\end{equation}

The generalization to n-phonon is straightforward

\begin{eqnarray}
\overline{\sigma ^{n}} &=&\sigma _{h}^{\left( n\right) }+\sigma _{c}^{\left(
n\right) }\stackrel{n}{%
\mathrel{\mathop{\sum }\limits_{k=1}}%
}\frac{(n-k)!}{n!}\frac{n-k}{n+k}%
{n \choose k}%
\left( \frac{\Gamma _{d_{1}}\tau _{c}\left( E\right) }{\hslash }\right) ^{k}
\nonumber \\
&\equiv &\sigma _{c}^{\left( n\right) }+\sum_{k=1}^{n}\sigma _{fl}^{\left(
n\right) }\left( k\right) \text{ .}
\end{eqnarray}

Note that $\sigma _{c}^{(n)}$ is the cross-section for the excitation of
n-phonon states that proceeds though $n-1,n-2,...$, phonon states which have
finite life times (width). In this respect we are generalizing the concept
of harmonic cross-section. The consideration of the width of the
intermediate states and the fluctuations that result from the decay of these
states go hand in hand.

Our analytical formula for the average cross-section is quite reasonable
when compared to numerical solution of the evolution equation for the
density matrix of the system reported in ref. [3] if the correlation time is
identified with the collision time at the grazing impact parameter, $\tau
_{c}=b_{o}/\gamma v$, Fig. 1. In view of this we are confident in applying
our theory to calculate other observables such as the width of the final
channel.

Taking a Breit-Wigner form for the spectrum of each of the component in (12)
and evaluating this at the peak, we can define an effective width as

\begin{equation}
\frac{\sigma _{c}^{\left( n\right) }}{n\Gamma _{d_{1}}}+\stackrel{n-1}{%
\mathrel{\mathop{\sum }\limits_{k=1}}%
}\frac{\sigma _{fl}^{\left( n\right) }\left( k\right) }{\left( n-k\right)
\Gamma _{d_{1}}}=\frac{\sigma ^{(n)}}{\Gamma _{eff}^{\left( n\right) }}\text{
,}
\end{equation}

\noindent or

\begin{equation}
\Gamma _{eff}^{\left( n\right) }\left( E\right) =\left( \frac{\sigma
_{h}^{\left( n\right) }+\Sigma _{k}\sigma _{fl}^{\left( n\right) }\left(
k\right) }{\sigma _{c}^{\left( n\right) }+\stackrel{n-1}{%
\mathrel{\mathop{\Sigma }\limits_{k=1}}%
}\frac{n}{n-k}\sigma _{fl}^{\left( n\right) }\left( k\right) }\right)
n\Gamma _{d_{1}}=\frac{1}{1+\frac{\stackrel{n-1}{%
\mathrel{\mathop{\Sigma }\limits_{k=1}}%
}\text{ }\frac{k}{n-k}\text{ }\sigma _{fl}^{\left( n\right) }\left( k\right) 
}{\sigma ^{\left( n\right) }}}\left( n\Gamma _{d_{1}}\right) \text{ .}
\end{equation}

Eq. (14) for $\Gamma _{eff}^{\left( n\right) }\left( E\right) $ shows that
the effective width attains the harmonic value $n\Gamma _{d_{1}}$ at high
energies. At low bombarding energy the value can be significantly smaller
than $n\Gamma _{d_{1}}$, owing to the Brink-Axel reduction factor $\left( 1+%
\frac{\stackrel{n}{%
\mathrel{\mathop{\Sigma }\limits_{k=1}}%
}\text{ }\frac{k}{n-k}\text{ }\sigma _{fl}^{\left( n\right) }\left( k\right) 
}{\sigma ^{\left( n\right) }}\right) ^{-1}$. This is demonstrated in Fig. 2.
The B-A mechanism thus supplies a correlation between the experimental
effective width and the corresponding cross-section. This is an important
result which further helps shed light on the excitation mechanism. For two
phonons we have $\frac{\sigma ^{\left( 2\right) }}{\sigma _{c}^{\left(
2\right) }}=\frac{1+\frac{\Gamma _{eff}}{2\Gamma _{1}}}{2%
{\Gamma _{eff} \overwithdelims() 2\Gamma _{1}}%
}$. This relation is independent of the system. In figure (3) we exhibit the
available data taken from ref. [6]. In presenting the data, we have changed
the variable $\frac{\sigma _{\exp .}^{\left( 2\right) }}{\sigma
_{harm.}^{\left( 2\right) }}$\ into $\frac{\sigma _{\exp .}^{\left( 2\right)
}}{\sigma _{c}^{\left( 2\right) }}$, with $\sigma _{harm.}^{\left( 2\right)
} $being the value of $\sigma _{c}^{\left( 2\right) }$ when the width of the
intermediate one-phonon state is set equal to zero.\ Although it is claimed
that $\Gamma _{eff}^{\left( 2\right) }=\sqrt{2}\Gamma _{1}$, our result does
not exclude a more subtle connection between $\Gamma _{eff}^{\left( 2\right)
}$ and $\Gamma _{1}$.

The above discussion suggests defining an effective phonon number $
\begin{array}{l}
n_{eff}^{\left( E\right) }\equiv \frac{n}{\frac{1+\stackrel{n-1}{%
\mathrel{\mathop{\Sigma }\limits_{k=1}}%
}\frac{k}{n-k}\sigma _{fl}^{\left( n\right) }\left( k\right) }{\sigma
^{\left( n\right) }}}
\end{array}
$ which is smaller than $n$. This gives an immediate qualitative explanation
for the enhancement in the cross-section too. Since $n_{eff}<n$, the energy $%
E_{n_{eff}}<nE_{1}$. Accordingly, if one were to use damped anharmonic\
oscillator model with the following sequence of energies, $0\rightarrow
E_{1}\rightarrow 2E_{1}-\frac{\Gamma \tau }{1+\Gamma \tau }E_{1}\rightarrow
3E_{1}-\frac{\Gamma \tau +2\left( \Gamma \tau \right) ^{2}}{1+\Gamma \tau
+\left( \Gamma \tau \right) ^{2}}E_{1}$, ..., one would obtain larger
cross-sections $\sigma ^{\left( 2\right) }$; $\sigma ^{\left( 3\right) }$,
... when compared to $\sigma _{c}^{\left( 2\right) }$, $\sigma _{c}^{\left(
3\right) }...$. In a way, this supplies a simple connection between the
direct + fluctuation model and the anharmonic models [7-9]. In Fig. 4 we
show the result of a calculation following this line.\ The multiphonon
cross-section calculated for the effective damped anharmonic oscillator
(long-dashed line), whose spectrum is given by $E_{n_{eff}}=n_{eff}$ $E_{1}$%
. The semiclassical coupled channels model [4b] is used to generate this
result. The widths of the intermediate states are just the ``harmonic'' ones 
$\Gamma _{n}=n\Gamma _{1}$. Also shown is the cross-section from Ref. [3]
(solid line) and the simple undamped harmonic oscillator cross-section also
calculated according to [4b] (small dashed line). The results shown are for
the system $^{208}$Pb+$^{208}$Pb. The agreement of our damped anharmonic
oscillator, coupled channels calculation with those of Ref. [3], which is
based on the solution of an evolution equation for the excitation
probability with loss and gain terms, is excellent.

In conclusion, the direct$+$fluctuation model (Brink-Axel mechanism)
supplies a natural framework to discuss the cross-section enhancement and
the width reduction of the multiphonon states. The enhancement factor of the
cross-section $\frac{\sigma ^{\left( n\right) }}{\sigma _{c}^{\left(
n\right) }}$ and the width reduction $\frac{\Gamma _{eff}^{\left( n\right) }%
}{n\Gamma _{d_{1}}}$ can be co-related by the effective number of phonons.
This concept should be quite useful in future work on multiphonon physics.
Further, our results here should also be relevant to other multistep
reaction phenomena such as preequilibrium processes [10].

\newpage

\centerline {\bf FIGURE CAPTIONS}

\begin{description}
\item[Fig. 1:]  The quantity $\Gamma ^{\downarrow }\tau _{c}$ as a function
of the bombarding energy for the $^{208}$Pb + $^{208}$Pb system.\bigskip 

\item[Fig. 2:]  The effective observed width $\Gamma _{eff}^{\left( n\right)
}$, Eq. (14) vs. the bombarding energy for the $^{208}$Pb+$^{208}$Pb system.
Clearly $\Gamma _{eff}^{\left( 1\right) }\equiv \Gamma \left( d_{1}\right) $%
, and it is set equal to 4.0MeV. See text for details.\bigskip 

\item[Fig. 3:]  The effective width reduction vs. the cross-section
enhancement. The data are from Ref. [6]. The ``data'' for the cross-section
ratios is $\frac{\sigma _{\exp .}^{\left( 2\right) }}{\sigma _{c}^{\left(
2\right) }}$ which is just the values of Ref. [6], $\frac{\sigma _{\exp
.}^{\left( 2\right) }}{\sigma _{harm.}^{\left( 2\right) }}$, multiplied by $%
\frac{\sigma _{c}^{\left( 2\right) }\left( \Gamma _{d_{1}}=0\right) }{\sigma
_{c}^{\left( 2\right) }}$, since $\sigma _{harm.}^{\left( 2\right) }\equiv
\sigma _{c}^{\left( 2\right) }\left( \Gamma _{d_{1}}=0\right) $. See text
for details.\bigskip 

\item[Fig. 4:]  The multiphonon cross-section calculated for the effective
damped anharmonic oscillator (long-dashed line), whose spectrum is given by $%
E_{n_{eff}}=n_{eff}$ $E_{1}$. The semiclassical coupled channels model [4b]
is used to generate this result. The widths of the intermediate states are
just the ``harmonic'' ones $\Gamma _{n}=n\Gamma _{1}$. Also shown is the
cross-section from Ref. [3] (solid line) and the simple undamped harmonic
oscillator cross-section also calculated according to [4b] (small dashed
line). The results shown are for the system $^{208}$Pb+$^{208}$Pb. See text
for details.
\end{description}

\end{document}